\documentclass[prl,twocolumn,showpacs,a4paper]{revtex4}
\usepackage{bm,graphicx,amsmath}

\newcommand{\ket}[1]{\left|#1\right\rangle}

\newcommand{\melc}[3]{\left\langle#1\left|#2\right|#3\right\rangle}

\begin{document}

\title{Quantum Phase Gate Operation Based on Nonlinear Optics: Full Quantum Analysis}
\author{C. Ottaviani, S. Rebi\'{c}, D. Vitali and P. Tombesi}
\affiliation{Dipartimento di Fisica, Universit\`{a} di Camerino, I-62032 Camerino
(MC), Italy}

\begin{abstract}
We present a full quantum treatment of a five-level atomic system coupled to two quantum and two classical light fields.
The two quantum fields undergo a cross-phase modulation induced by electromagnetically induced transparency.
The performance of this configuration as a two-qubit quantum phase gate for travelling single-photons is examined.
A trade-off between the size of the conditional phase shift and the fidelity of the gate is found.
Nonetheless, a satisfactory gate performance is still found to be possible in the transient regime, corresponding to a fast gate operation.
\end{abstract}

\maketitle

Single photons are natural candidates for the implementation of quantum information processing systems \cite{chuang95}. This is due to the photon's
robustness against decoherence and the availability of single-qubit operations.
However, it is difficult to realize the necessary two-qubit operations since the interaction
between photons is very small. A possible solution is the enhancement of photon-photon interaction
either in cavity QED configurations \cite{turch} or in dense atomic media exhibiting electromagnetically induced transparency (EIT) \cite{eit}.
In this latter case, optical nonlinearities can be produced when EIT is disturbed, either by introducing additional energy level(s)~\cite{Schmidt96,Wang01},
or by mismatching the probe and control field frequencies~\cite{Grangier98,Matsko03}.

\begin{figure}[t]
\includegraphics[scale=0.8]{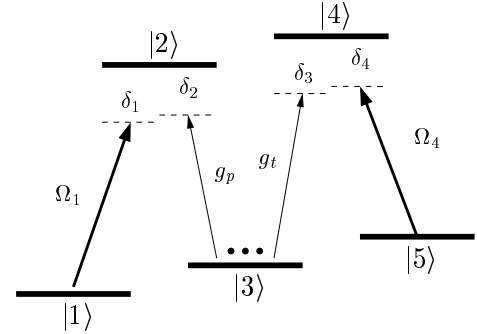}
    \caption{Energy levels of the ``M''-scheme. $\Omega_j$ are the Rabi frequencies of classical fields, while $g_{p,t}$
    denote couplings of the quantized probe and trigger fields to their respective transitions. $\delta_j$ are the detuning
    of the fields from resonance.} \label{fig:levelscheme}
\end{figure}

In this letter, we address the feasibility of EIT-based systems for the implementation of a two-qubit quantum phase gate (QPG) for travelling
single photons~\cite{Lukin00,Ottaviani03,Masalas04}, by means of a \emph{full quantum} treatment of the system dynamics.
In a QPG, one qubit gets a phase conditional to the other qubit state according to the transformation~\cite{Lloyd95,NielsenChuang} $|i\rangle _{1}|j\rangle _{2}
\rightarrow \exp\left\{i \phi_{ij} \right\}|i\rangle _{1}|j\rangle _{2} $ where $\{i,j\}=0,1$ denote the logical qubit bases. This gate is universal
when the conditional phase shift (CPS)
\begin{equation}\label{eq:def_cps}
\phi = \phi_{11} + \phi_{00} -\phi_{10} -\phi_{01},
\end{equation}
is nonzero, and it is equivalent to a CNOT gate up to local unitary transformations when $\phi=\pi$ \cite{Lloyd95,NielsenChuang}.
The existing literature focused only on the evaluation of the CPS and on the best conditions for achieving
$\phi=\pi$~\cite{Lukin00,Ottaviani03,Masalas04}, while the gate fidelity, which is the main quantity
for estimating the efficiency of a gate, has been never evaluated. In this letter we calculate \emph{both the fidelity and the CPS} of the QPG,
enabling us to discover a general \emph{trade--off} between a large CPS and a gate fidelity close to one, hindering the QPG operation.
However, we show that this trade-off can be bypassed in the transient regime, which has never been considered before in
EIT situations, still allowing a satisfactory gate performance.

The qubits are given by polarized single--photon wave packets with different frequencies, and the phase shifts $\phi_{ij}$
are generated when these two pulses cross an atomic ensemble in a five-level ``M'' configuration (see  Fig.~\ref{fig:levelscheme}).
The population is assumed to be initially in the ground state $\ket{3}$. From this ground state, it could be excited by either the
single--photon {\em probe} field, coupling to transition $\ket{3} \leftrightarrow \ket{2}$, or by the single--photon {\em trigger}
field, coupling to transition $\ket{3} \leftrightarrow \ket{4}$. If the five levels are Zeeman sub-levels of an alkali atom, and
both pulses have a sufficiently narrow bandwidth, the Zeeman splittings can be chosen so that the atomic medium is coupled only
to a given circular polarization of either the probe or trigger field, while it is transparent for the orthogonally polarized mode,
which crosses the gas undisturbed \cite{Ottaviani03}. In this way, the logical basis for each qubit practically coincides with the
two lowest Fock states of the mode with the ``right'' polarization, $|0_j\rangle$ and $|1_j\rangle$ ($j=p,t$).

When the probe (trigger) is on two--photon resonance with the classical pump field with Rabi frequency $\Omega_1\ (\Omega_4)$,
i.e. $\delta_1=\delta_2\ (\delta_3=\delta_4)$ (see Fig.~\ref{fig:levelscheme} for a definition of the detunings), the system exhibits
EIT for probe and trigger simultaneously. In fact, the scheme can be seen as formed by two adjacent $\Lambda$ systems, perfectly symmetric
between probe and trigger. A nonzero CPS occurs whenever a nonlinear cross-phase modulation (XPM) between probe and
trigger is present. This cross-Kerr interaction takes place if the two-photon resonance condition is violated. For small frequency
mismatch $\epsilon_{12}=\delta_1-\delta_2$ and $\epsilon_{34}=\delta_3-\delta_4$ (both chosen to be within the EIT window),
absorption remains negligible and the cross-Kerr interaction between probe and trigger photons may be strong. The consequent
CPS may become large, of the order of $\pi$, if the probe and trigger pulse simultaneously cross the atomic medium and
interact for sufficient time. This is achieved when the group velocities of the two pulses are small and equal (to $v_g$, see Refs.~\cite{Lukin00,Ottaviani03}),
so that the interaction time is given by $t_{int}=L/v_g$, $L$ being the length of the gas cell \cite{cigar}. The inherent
{\em symmetry} of this scheme guarantees perfect group velocity matching whenever $\delta_1=\delta_4$, $\delta_2=\delta_3$
and $g_{p}/\Omega_{1}=g_{t}/\Omega_{4}$, where $g_j=\mu_j \sqrt{\omega_j/2\hbar \epsilon_0 V}$ ($j=p,t$) is the coupling
constant between the probe (trigger) quantum mode with frequency $\omega_j$ and the corresponding transition with electric
dipole moment $\mu_j$. These  features are shared by all the proposals for an EIT-based, nonlinear
two-qubit quantum gate \cite{Lukin00,Ottaviani03}. They essentially differ only in the way in which group velocity matching
is achieved.

The scope of this paper is to find the ultimate \emph{physical} limits imposed on QPG operations in systems with EIT-based optical nonlinearities.
To this end, we neglect all the possible technical limitations and experimental imperfections.
First, we assume perfect spatial mode matching between the input single-photon pulses entering the gas cell and the optical modes
excited by the driven atomic medium, and which are determined by the geometrical properties of the gas cell and of the pump beams \cite{Duan02}.
This allows us to describe the probe and trigger fields in terms of \emph{single} travelling optical modes, with annihilation operators $\hat{a}_{p,t}$.
Next, we assume that the pulses are tailored in such a way that they simultaneously enter the gas cell and completely overlap with it
during the interaction. This means that their length (compressed due to group velocity reduction) is of the order of the cell length $L$
and their beam waist is of the order of the cell radius. In this way, the two pulses interact with \emph{all} $N_a$ atoms in the cell,
and moreover one can ignore spatial aspects of pulse propagation. With these assumptions, and neglecting dipole-dipole interactions,
the interaction picture Hamiltonian may be written as
\begin{eqnarray}\label{eq:coll_ham}
&& H = \hbar\epsilon_{12}\hat{S}_{11}+\hbar\delta_{2}\hat{S}_{22} +\hbar\delta_{3}\hat{S}_{44}+\hbar\epsilon_{34}\hat{S}_{55}  \\
&&+\hbar\Omega_{1}\sqrt{N_a}\left(\hat{S}_{21}+\hat{S}_{12}\right)+\hbar g_{p}\sqrt{N_{a}}\left(\hat{a}_{p}\hat{S}_{23}+\hat{S}_{32}\hat{a}_{p}^{\dagger}\right) \nonumber\\
&&+\hbar g_{t}\sqrt{N_{a}}\left(\hat{a}_{t}\hat{S}_{43}+\hat{S}_{34}\hat{a}_{t}^{\dagger}\right)+ \hbar\Omega_{4}\sqrt{N_a}\left(\hat{S}_{45}+\hat{S}_{54}\right),
\nonumber
\end{eqnarray}
where we have defined the collective atomic operators
$ \hat{S}_{kl}=\sum_{i=1}^{N_a}\sigma_{kl}^i/\sqrt{N_{a}}$, $k\neq l=1,\ldots,5$,
and $ \hat{S}_{kk}=\sum_i\sigma_{kk}^i $, with $\sigma_{kl}^i\equiv |k\rangle _i \langle l |$ referring to the $i$th atom. Since the initial state
\begin{eqnarray}
&& |\psi_{in}\rangle = \bigotimes_{i=1}^{N_a} \ket{3}_i \otimes \left(c_{00}|0_p\rangle \otimes |0_t\rangle+c_{01}|0_p\rangle \otimes|1_t\rangle \right.
\nonumber \\
&& \left. +c_{10}|1_p\rangle \otimes |0_t\rangle
+c_{11}|1_p\rangle \otimes |1_t\rangle \right) \label{iniz}
\end{eqnarray}
contains at most two excitations, the time evolution driven by Eq.~(\ref{eq:coll_ham}) is simple and restricted to a finite-dimensional
Hilbert space involving few symmetric collective atomic states. In fact, each component of the initial state of Eq.~(\ref{iniz})
evolves independently in a different subspace. Defining $|e_3^{(n_p,n_t)}\rangle = \bigotimes_{i=1}^{N_a} \ket{3}_i |n_p\rangle \otimes  |n_t\rangle$,
the component with no photon in Eq.~(\ref{iniz}), $|e_3^{(0,0)}\rangle$, is an eigenstate of $H$ and does not evolve.
The component $|e_3^{(0,1)}\rangle$ evolves in a three-dimensional
Hilbert space spanned also by the two states $|e_4^{(0,0)}\rangle $ and $|e_5^{(0,0)}\rangle$, where we have defined, for $r = 1,2,4,5$,
the symmetric collective states
\begin{equation}\label{eq:gen_state}
|e_r^{(n_p,n_t)}\rangle  = \frac{1}{\sqrt{N_{a}}}\sum_{i=1}^{N_{a}}\ket{3_{1},3_{2},\dots,r_{i},\dots,3_{N_{a}}} \otimes\ket{n_{p}}\otimes\ket{n_{t}}.
\end{equation}
In a similar fashion, the component with only one probe photon, $|e_3^{(1,0)}\rangle$,  evolves in a three-dimensional Hilbert space
spanned also by the two states $|e_1^{(0,0)}\rangle $ and $|e_2^{(0,0)}\rangle $.  Finally, the component $|e_3^{(1,1)}\rangle$ evolves in the
five dimensional subspace spanned also by the four collective states $|e_1^{(0,1)}\rangle $, $|e_2^{(0,1)}\rangle $, $|e_4^{(1,0)}\rangle $,
and $|e_5^{(1,0)}\rangle$. What is relevant is that the dynamics remain simple and restricted within a finite-dimensional Hilbert space even when we include
spontaneous emission, so that time evolution is described by the following master equation for the system density matrix $\rho$,
\begin{equation} \label{eq:mastercoll}
\dot{\rho} = -\frac{i}{\hbar}\left[ H,\, \rho\right]
+ \sum_{kl}\frac{\gamma_{kl}}{2}\sum_{j=1}^{N_a}\left( 2\sigma_{kl}^j\rho\sigma_{kl}^{j \dagger} - \sigma_{kl}^{j \dagger}\sigma_{kl}^j\rho
- \rho\sigma_{kl}^{j \dagger}\sigma_{kl}^j \right),
\end{equation}
where $\gamma_{kl}$ denotes the decay rate from the excited states $l=2,4$ to the ground  states $k=1,3,5$ \cite{dephasing}.
Spontaneous emission seems to complicate the system dynamics. However, the Hamiltonian evolution involves
only the \emph{singly excited} symmetric atomic states of Eq.~(\ref{eq:gen_state}). This means that these collective states decay
with a rate equal to the single-atom decay rate $\gamma_{kl}$, and that spontaneous emission involves only a restricted number
of additional collective atomic states in the dynamics. To state it in an equivalent way, the atomic medium behaves
as an effective \emph{single} $5$-level atom, with a collectively enhanced coupling
with the optical modes $g_j\sqrt{N_a}$, but with the same single-atom decay rates $\gamma_{kl}$, Rabi frequencies $\Omega_i$, and detunings $\delta_i$
(see also Ref.~\cite{Duan01}).

Spontaneous emission causes the four independent Hilbert subspaces corresponding to the four initial state components to become coupled.
Moreover, the joint effect of the ``cross'' decay channels $|4\rangle \to |1\rangle $ and $|2 \rangle \to |5\rangle $ together with the Hamiltonian dynamics
couples the above-mentioned collective states
with six new states, $|e_1^{(1,0)}\rangle$, $|e_2^{(1,0)}\rangle $, $|e_3^{(2,0)}\rangle $ (populated if $\gamma_{41}\neq 0$), and
$|e_5^{(0,1)}\rangle $, $|e_4^{(0,1)}\rangle $,  $|e_3^{(0,2)}\rangle $ (populated if $\gamma_{25}\neq 0$). Therefore Eq.~(\ref{eq:mastercoll})
actually describes dynamics in a Hilbert space of dimension $18$, which we have numerically solved in order to establish the performance of the QPG.

This analysis allows us to fully characterize the QPG operation, by calculating \emph{both} the CPS $\phi$ of Eq.~(\ref{eq:def_cps})
and the fidelity of the gate, at variance with former treatments \cite{Lukin00,Ottaviani03,Masalas04}.
The accumulated CPS as a function of $t_{int}$ is obtained by using the fact that the
phase shifts $\phi_{ij}$ of Eq.~(\ref{eq:def_cps}) are given by combinations of the phases of the off-diagonal matrix elements (in the Fock basis)
of the reduced density matrix of the probe and trigger modes, $\rho_f(t_{int})$. The gate fidelity is given by \cite{NielsenChuang}
\begin{equation} \label{fid}
{\mathcal F}(t_{int}) = \sqrt{\overline{\melc{\psi_{id}(t_{int})}{\rho_{f}(t_{int})}{\psi_{id}(t_{int})}}},
\end{equation}
where $\ket{\psi_{id}(t_{int})} = c_{00}\exp\{i\phi_{00}(t_{int})\} |0_p,0_t\rangle+c_{01}\exp\{i\phi_{01}(t_{int})\}|0_p,1_t\rangle + c_{10}\exp\{i\phi_{10}(t_{int})\}
 |1_p,0_t\rangle
+c_{11}\exp\{i\phi_{11}(t_{int})\}|1_p,1_t\rangle$ is the ideally evolved state from the initial condition (\ref{iniz}), with phases $\phi_{ij}(t_{int})$
evaluated from $\rho_f(t_{int})$ as discussed above. The overbar denotes the average over all initial states (i.e., over the $c_{ij}$, see Ref.~\cite{Poyatos97}).
The above fidelity characterizes the performance of the QPG as a deterministic gate. However, one could also consider the QPG as a \emph{probabilistic} gate,
whose operation is considered only when the number of output photons is equal to the number of input photons. The performance of this probabilistic QPG could
be experimentally studied by performing a conditional  detection of the phase shifts, and it is characterized by the \emph{conditional} fidelity
${\mathcal F}^c(t_{int})$, similar to that of Eq.~(\ref{fid}), but with $\rho_f(t_{int})$ replaced by $\rho_f^c(t_{int})={\rm Tr}_{atom}\{|\psi_{nj}(t_{int})
\rangle \langle \psi_{nj}(t_{int})|\}/\langle \psi_{nj}(t_{int})|\psi_{nj}(t_{int})\rangle $, where $|\psi_{nj}(t_{int})\rangle$ is the (non-normalized)
evolved atom-field state conditioned to the detection of no quantum jumps \cite{Carmichael93}, i.e., of no spontaneous emission.

The conditional fidelity is always larger than the unconditional one, but they become equal (and both approach $1$) for an ideal QPG in which the number
of photons is conserved and all the atoms remain in state $|3\rangle$. This ideal condition is verified in the limit of large detunings $\delta_j \gg \gamma_{kj}$
(to significantly suppress spontaneous emission) and very small couplings $g_j\sqrt{N_a}\ll \Omega_j$. In this limit, each component of the
initial state of Eq.~(\ref{iniz}) practically coincides with the dark state of the four independent Hamiltonian dynamics discussed above.
The four phase shifts $\phi_{ij}$ can be evaluated as a fourth-order perturbation expansion of the corresponding eigenvalue, multiplied by
$t_{int}$, obtaining the following CPS
\begin{eqnarray}\label{eq:phipert}
&& \phi = \frac{g_{p}^{2}g_{t}^{2}N_a^2 t_{int}}{(\epsilon_{34}\delta_{3} - \Omega_{4}^{2})(\epsilon_{12}\delta_{1} - \Omega_{1}^{2})} \\
&&\times \left[\frac{\epsilon_{34}(\epsilon_{12}^{2}+\Omega_{1}^{2})}{(\epsilon_{12}\delta_{1} - \Omega_{1}^{2})} + \frac{\epsilon_{12}(\epsilon_{34}^{2}
+ \Omega_{4}^{2})}{(\epsilon_{34}\delta_{3} - \Omega_{4}^{2})}\right]. \nonumber
\end{eqnarray}
This prediction is verified by the numerical solution of Eq.~(\ref{eq:mastercoll}) in the limit of large detunings and small couplings.
However the resulting CPS is too small, even for very long interaction times (i.e., long gas cells): for example, for $g_{p,t}\sqrt{N_a}=0.5$ MHz,
$\epsilon_{12,34}=1.9$ MHz, $\Omega_{1,4}=65$ MHz and $\delta_{1,3}=1.9$ GHz, we obtain a tiny CPS of only $3\times 10^{-4}$ radians when
$t_{int}=10^{-4}$ s. This is not surprising because this limit corresponds to a dispersive regime far from EIT, and one has to explore the
non-perturbative regime of larger couplings in order to exploit EIT and achieve a satisfactory QPG operation.

\begin{figure}[t]
\includegraphics[scale=0.85]{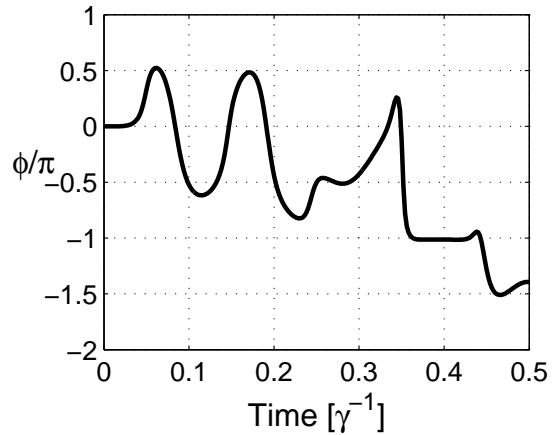}
    \caption{Conditional phase shift $\phi$ versus the interaction time. See text for details.} \label{fig:cps}
\end{figure}

\begin{figure}[t]
\includegraphics[scale=0.85]{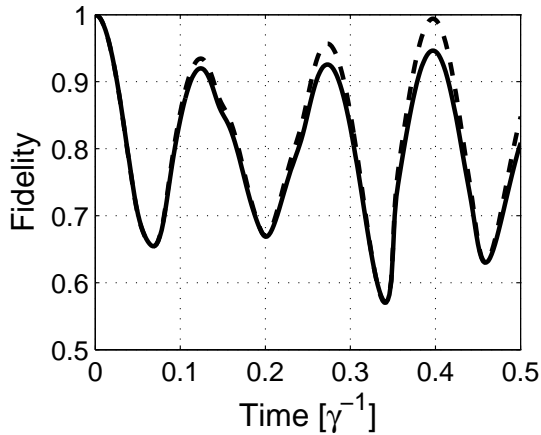}
    \caption{Fidelity of the QPG operation. Deterministic fidelity (solid) and conditional fidelity (dashed) are shown. See text for details.} \label{fig:fidelity}
\end{figure}

We have found good QPG performance for the following parameters, corresponding to a gas cell of $N_a \simeq 10^8$ $^{87}$Rb atoms: $\gamma_{kl}=\gamma =
2\pi \times 6$ MHz, $\delta_1 = \delta_3 = 15\gamma$, $\epsilon_{12} = \epsilon_{34} = 0.01\gamma$, $g_p = g_t = 0.0022\gamma$, $\Omega_1 = \Omega_4 =
4\gamma$.  The results are shown in Figs.~\ref{fig:cps} and~\ref{fig:fidelity}, where we see that a CPS of $\sim \pi$ radians is obtained in the
transient regime for $t_{int}\approx 0.4/\gamma \sim 10$ ns, corresponding to a \emph{fast operation} of the gate.
At the same interaction time, the unconditional gate fidelity (Fig.~\ref{fig:fidelity},
full line) is about $94\%$, while the conditional gate fidelity reaches the value of $99\%$ (Fig.~\ref{fig:fidelity}, dashed line), in correspondence
with a success probability of the gate equal to $0.94$. The probe and trigger group velocity is $v_g \simeq 3 \times 10^6$ ms$^{-1}$,
yielding a gas cell length $L = v_g t_{int} \simeq 3.1$ cm. The value of $g_j$ yields an interaction volume $V \simeq 2 \cdot 10^{-3}$ cm$^3$,
corresponding to a gas cell diameter of about $330$ $\mu$m and to an atomic density $N_a/V \simeq 5 \cdot 10^{10}$ cm$^{-3}$.

EIT is a stationary phenomenon, while the above results are obtained in the transient regime where $\gamma t_{int} < 1$. However we can attribute these
results to a sort of ``non-stationary'', EIT process. This is suggested by the reduction of $v_g$ (by a factor $\simeq 100$), which has been estimated by
evaluating the ``instantaneous'' susceptibility from the reduced atomic density matrix given by Eq.~(\ref{eq:mastercoll}) and then averaging over the time
interval between $0$ and $t_{int}$. This ``non-stationary'' $v_g$ is one order of magnitude smaller than the conventional $v_g$ obtained from the steady-state
susceptibility corresponding to the above parameters. The presence of a moderate EIT
process is also confirmed by the fact that in a numerical study of the three-level ladder atomic scheme, yielding XPM without EIT \cite{Schmidt96},
we have found a slower accumulation of the CPS and a smaller conditional fidelity ($\sim 78\%$) for a corresponding set of parameters.

Our study of Eq.~(\ref{eq:mastercoll}) also shows that it is not possible to achieve a comparable QPG performance in the steady-state regime
$\gamma t_{int} \gg 1$. In fact, we have found at best a CPS of $\pi$ in correspondence with fidelities ${\mathcal F}(t_{int})$ and ${\mathcal F}^c(t_{int})$
equal to $77\%$ and $83\%$, respectively. This is due to the general presence of a trade-off between the size of the CPS and of the gate fidelity.
In fact, we have seen that both gate fidelities approach $1$ in the small perturbation limit, but with a CPS which becomes appreciable only for
unrealistically long gas cells. A larger CPS requires a larger ratio $g_j\sqrt{N_a}/\Omega_j$. This condition however increases the population
of atomic states $\ket{1}$ and $\ket{5}$ at the expense of the initial atomic state $\ket{3}$, unavoidably decreasing the gate fidelity.
Similar conclusions hold for other options, such as increased detunings $\delta_j$, or adjusting two-photon detunings $\epsilon_{ij}$.
This trade-off is present also at large ratios $g_j\sqrt{N_a}/\Omega_j$
in the transient regime, where, however, it may be
less effective. In fact, in this case one has significant oscillations of the atomic populations,
but it is possible to find appropriate interaction times $t_{int}$ at which high fidelities are achieved (see Fig.~\ref{fig:fidelity}), simultaneously
with a CPS of about $\pi$.

In conclusion, our study shows that the implementation of efficient EIT-based nonlinear two-qubit gates
for travelling single-photons is possible. In fact, even if there is
a trade-off between the size of the CPS and the fidelity of the gate in the stationary regime, it is possible to have a satisfactory gate performance
in the transient regime, where a fast gate operation and fidelities equal to $0.99$ are achievable. The
experimental realization might be challenging, but the implementation of this quasi-deterministic two-qubit gate would be extremely useful,
not only for quantum computation, but also for quantum communication purposes: for example, a QPG allows a complete Bell-state discrimination
for single-photon polarization qubits \cite{Vitali00}.
We expect that these considerations apply to all EIT-based crossed-Kerr schemes \cite{Lukin00,Ottaviani03},
regardless of the specific level scheme considered. Finally, we note that our analysis does not apply to situations where the nonlinearity comes from independent
processes such as atomic collisions or dipole-dipole interactions~\cite{Masalas04}.

We acknowledge enlightening discussions with G. Di Giuseppe.


\begin{thebibliography}{99}
\bibitem{cigar}We assume a cylindrical atomic medium with the pump beams propagating along its axis,
collinear with the two quantum fields in order to avoid Doppler broadening.
\bibitem{dephasing}We neglect atomic dephasing,
which is expected to give negligible effects in the transient regime and dilute gas situation we consider in the paper.
\end{thebibliography}
\end{document}